\documentclass[11pt]{article} \hoffset=-15mm \voffset=-10mm
\usepackage{graphicx,amsmath,amssymb,epsf} \usepackage{latexsym,bm,
  slashed} \usepackage{xcolor} \definecolor{dark}{rgb}{0.10,0.2,0.3}
\definecolor{magenta}{rgb}{0.7,0.1,0.3}
\definecolor{purpure}{rgb}{0.5,0.15,0.3}
\usepackage[font=small,format=plain,labelfont=bf,up,textfont=it,up]{caption}
\usepackage{hyperref, cite, array, booktabs} \hypersetup{colorlinks, citecolor=blue,
  filecolor=blue, linkcolor=magenta,
  urlcolor=purpure,hyperfootnotes=true,pdftex}

 \title{\bf  \Large Exclusive 
$J/\Psi$ and $\Psi(2s)$ photo-production as a probe of QCD low $x$ evolution equations      
   } \author{Martin~Hentschinski and  Emilio~Padr\'on~Molina \\ \\
Departamento de Actuaria, F\'isica y Matem\'aticas,
Universidad de las Americas Puebla, \\ Santa Catarina Martir, 72820 Puebla, Mexico
}

\begin{document}

\maketitle
\begin{abstract}
  We investigate photo-production of vector mesons $J/\Psi$ and
  $\Psi(2s)$, based on both HERA and LHC data, using 2 fits of
  unintegrated gluon distributions. The latter are subject to
  non-linear Balitsky-Kovchegov evolution (Kutak-Sapeta gluon; KS) and
  linear next-to-leading order Balitsky-Kuraev-Fadin-Lipatov evolution
  (Hentschinski-Sabio Vera-Salas; HSS gluon) respectively. Apart from
  extending previous studies to the case of radially excited
  charmonium $\Psi(2s)$, we further use an improved set of charmonium
  wave functions, provided in the literature, and give an estimate of
  the uncertainties associated with the energy dependence of the HSS
  gluon. While we observe that the difference between linear and
  non-linear evolution somehow diminishes and a clear distinction
  between both HSS and KS gluon is not possible using the currently
  available data set, we find that the differences between both gluon
  distributions are enhanced for the ratio of the photo-production
  cross-sections of $\Psi(2s)$ and $J/\Psi$ vector mesons.
 
\end{abstract}

\section{Introduction}
\label{sec:intro}

Due to its large center of mass energy, the Large Hadron Collider
(LHC) provides a unique opportunity to explore the dynamics of strong
interactions in the high energy or Regge limit. For a process with a
hard scale, which renders the strong
coupling constant $\alpha_s$ small, a study of the Regge limit is
possible using perturbative Quantum Chromodynamics (QCD). The
theoretical description is  provided through the
Balitsky-Fadin-Kuraev-Lipatov (BFKL) evolution, which achieves a
resummation of perturbative higher order corrections, which are enhanced by a
large logarithm in $x$ to all orders in the strong coupling at
leading (LL) \cite{Kuraev:1976ge, Lipatov:1976zz,
  Kuraev:1977fs, Balitsky:1978ic} and next-to-leading (NLL)
\cite{Fadin:1998py, Ciafaloni:1998gs} logarithmic accuracy. Here
$x = M^2/s$ where $M$ denotes the characteristic hard scale of the
process and $s$ the center of mass energy squared. The
perturbative high energy limit is then defined as $x \to 0$ at $M = $fixed. BFKL
evolution predicts a power-like rise of the proton structure function
$F_2$ with $1/x$, which is driven by the gluon distribution. While
this rise is seen in the data and can be described by BFKL evolution
\cite{Hentschinski:2012kr, Hentschinski:2013id, Kowalski:2017umu,
  Kowalski:2010ue}, it is known that it cannot continue down to
arbitrary small values of $x$. Instead, BFKL evolution will eventually
drive the proton into an over occupied system of gluons, which
eventually leads to the saturation of gluon densities
\cite{Gribov:1984tu}. Finding convincing and substantial evidence for
gluon saturation as well as for the transition into this region of QCD
phase space is still one of the open problems of QCD and at the core
of the physics program of the future Electron Ion Collider
\cite{Accardi:2012qut}. \\

A very useful observable to explore the gluon distribution  at the LHC  in this
region of interest is provided by exclusive
photo-production of vector mesons. The observable is somewhat
complementary to the bulk of studies currently undertaken
\cite{Bury:2020ndc, vanHameren:2019ysa, Kolbe:2020tlq, Mantysaari:2019hkq, Altinoluk:2020qet, Celiberto:2020rxb, Celiberto:2020tmb, Bolognino:2019yls, Celiberto:2018muu},
which attempt to resolve the hadronic final state in order to explore
the low $x$ gluon. In contrast to those studies, exclusive
photo-production of vector mesons allows for a direct observation of
the energy dependence of the photo-production cross-section which
directly translates into the $x$-dependence of the underlying gluon
distribution. In particular, if both HERA and LHC data are combined,
the probed region in $x$ extends over several orders of magnitude of
$x$, down to smallest values of $x = 4 \cdot 10^{-6}$. Photo-production of bound states of charm quarks, {\it
  i.e.}  $J/\Psi$ and $\Psi(2s)$ vector mesons, are then attractive
observables, since the charm mass provides a hard scale at the
border between soft and hard physics; the observable is therefore
expected to be particularly sensitive to the possible presence of a
semi-hard scale associated with the transition to the saturation
region, the so-called saturation scale. \\

In \cite{Bautista:2016xnp} it has been found that an unintegrated
gluon distribution subject to NLO BFKL evolution (the
Hentschinski-Salas-Sabio Vera gluon; HSS)
\cite{Hentschinski:2012kr, Hentschinski:2013id} is able to describe the
energy dependence of the photo-production cross-section of $J/\Psi$
and $\Upsilon$ vector mesons. In \cite{Garcia:2019tne}, this study has
been extended to the Kutak-Sapeta (KS) gluon \cite{Kutak:2012rf},
which is subject to non-linear Balitsky-Kovchegov (BK) evolution
\cite{Balitsky:1995ub, Kovchegov:1999yj}. While both gluon
distributions were able to describe the available data set, we found
that a certain perturbative expansion, which underlies the linear HSS
gluon, leads to an instability at highest values of the center of mass
energy $W$. While the instability can be removed through an improved
scale setting, the growth of the stabilized gluon distribution with
energy is too strong and linear evolution does no longer describe the
data-set. This observation was then interpreted as a first indication
for the transition towards saturated gluon densities.  Note that in
\cite{Flett:2020duk} is has been pointed out that this observation
does not indicate saturation of gluon densities, but mainly the need
for absorptive corrections (in the terminology of
\cite{Flett:2020duk}). We agree in principle with this observation:
the gluon does certainly not saturate at current values of the center
of mass energy; one merely finds signs for the slow down of the
power-like growth which points towards an increasing relevance of
non-linear terms in low $x$ QCD evolution equations. In other words,
the cross-section is about to enter the so-called transition region,
which separates the phase space region characterized by low and
saturated gluon densities respectively.  For a related study based on
a different implementation of BFKL evolution, see
\cite{Goncalves:2020ywm}, also \cite{Goncalves:2020feg}.
\\

In the present paper we extend the study of \cite{Garcia:2019tne}, to
the case of radially excited charm-anti charm states, {\it i.e.} the
$\Psi(2s)$ vector meson. As for photo-production of $J/\Psi$, the hard
scale is provided by the charm mass, placing us at the boundary
between soft and hard physics.  On the other hand, the dependence of
the light-cone wave function on the dipole size differs for $\Psi(2s)$
and $J/\Psi$.  We therefore expect to test with $\Psi(2s)$
photo-production a slightly different region in transverse momentum of
the unintegrated gluon distribution. To increase the precision of our
study we used instead of the previously implemented boosted Gaussian
model for the vector meson wave function \cite{Brodsky:1980vj,
  Nemchik:1994fp, Cox:2009ag}, a more refined description based on the
numerical solution of the Schr\"odinger equation for the charm-anti
charm state, provided in
\cite{Krelina:2018hmt, Cepila:2019skb}. \\

The outline of this paper is as follows: In Sec.~\ref{sec:setup} we
provide the technical details of our theoretical description, in
Sec.~\ref{sec:results} we present the results of our numerical study
and a comparison to data, while in Sec.~\ref{sec:conlc} we summarize
our results and draw our conclusions.

\section{Theoretical setup of our study}
\label{sec:setup}
\begin{figure}[t]
  \centering
   \includegraphics[width = .5\textwidth]{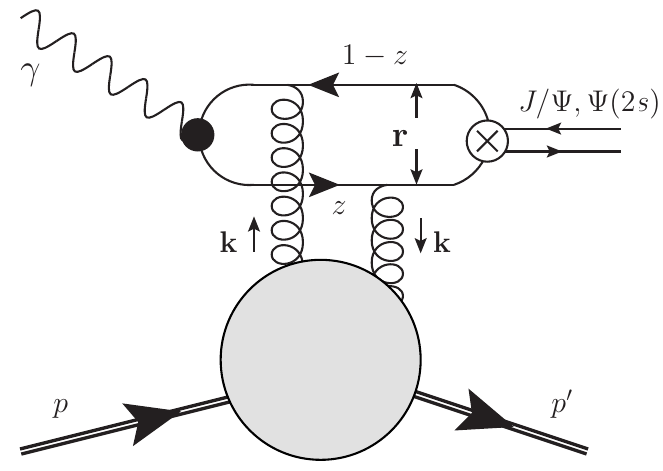}
  \caption{Exclusive photo-production of vector mesons $J/Psi$ and
    $\Psi(2s)$. For the quark-anti quark dipole we indicate photon momentum
    fractions $z$ and $1-z$ as well as the transverse separation ${\bm
    r}$. Finally ${\bm k}$ denotes the transverse momentum transmitted
from the unintegrated gluon distribution of the proton; the latter is indicated
through the gray blob.}
  \label{fig:reaction}
\end{figure}

\begin{figure}[t]
  \centering
  \parbox{.45\textwidth}{\includegraphics[width=.45\textwidth]{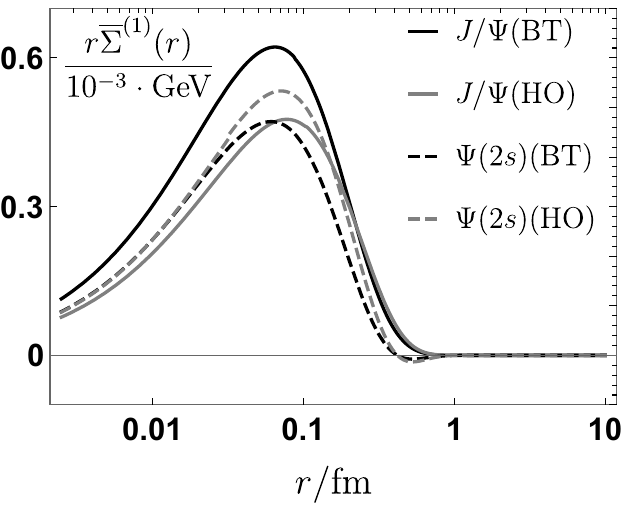}} \hspace{1cm}
 \parbox{.45\textwidth}{\includegraphics[width=.45\textwidth]{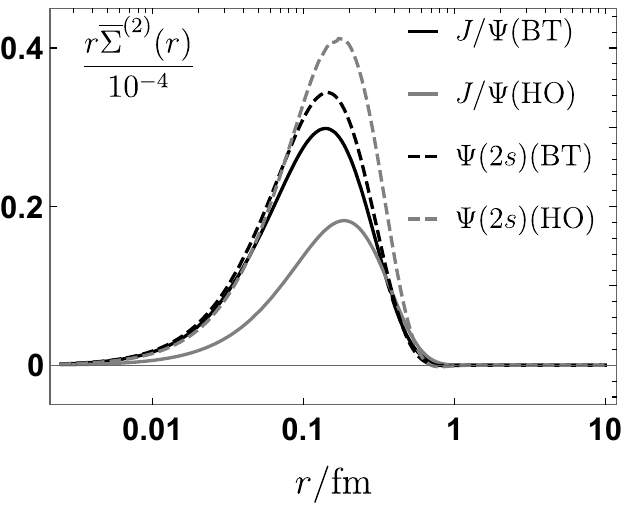}}
  \caption{The functions $\overline{\Sigma}_T^{(1)}$ (left) and $\overline{\Sigma}_T^{(2)}$(right) as defined in Eq.~\eqref{eq:sigma} and multiplied with a factor of $r$ for the Buchm\"uller-Tyle  and Harmonic Oscillator  potentials. }
  \label{fig:sigma}
\end{figure}

In the following we describe the framework on which our study is based, see also Fig.~\ref{fig:reaction}. We study the process
\begin{align}
  \label{eq:30}
 \gamma(q) + p(p) & \to V(q') + p(p') \, ,
\end{align}
where $V = J/\Psi, \psi(2S)$ while $\gamma$ denotes a quasi-real
photon with virtuality $Q \to 0$; $W^2 = (q + p)^2$ is the squared
center-of-mass energy of the $\gamma(q) + p(p)$
collision. With the momentum transfer $t = (q - q')^2$, the differential cross-section for
the exclusive photo-production of a vector meson can be written in the
following form
\begin{align}
  \label{eq:16}
  \frac{d \sigma}{d t} \left(\gamma p \to V p \right) 
& = 
\frac{1}{16 \pi} \left|\mathcal{A}_{T}^{\gamma p \to V p}(W^2, t) \right|^2 \, .
\end{align}
where $\mathcal{A}_T(W^2, t)$ denotes the scattering amplitude for the
reaction $\gamma p \to V p$ for color singlet exchange in the
$t$-channel, with an overall factor $W^2$ already extracted.  For a more detailed discussion see \cite{Bautista:2016xnp}.  In the following we  determine the total photo-production
cross-section, based on an inclusive gluon distribution. This is possible
following a two step procedure, frequently employed in the
literature: First one determines the differential cross-section at zero
momentum transfer $t=0$ (which can be expressed in terms of the
inclusive gluon distribution). In a second step  the
$t$-dependence is modeled, which then allows us to relate  the
differential cross-section at $t=0$ to the integrated
cross-section. In order to do so, we assume   an exponential drop-off
with $|t|$ of the differential cross-section, $\sigma \sim \exp\left[-|t| B_D(W)\right]$   with an energy dependent
$t$ slope parameter $B_D$,
\begin{align}
  \label{eq:18}
  B_D(W) & =\left[  b_0 + 4 \alpha' \ln \frac{W}{W_0} \right] \text{GeV}^{-2}.
\end{align}
The total cross-section for vector meson production is therefore
obtained as
\begin{align}
  \label{eq:16total}
 \sigma^{\gamma p \to V p}(W^2) & = \frac{1}{B_D(W)} \frac{d \sigma}{d t} \left(\gamma p \to V p \right)\bigg|_{t=0}
.
\end{align}
The  uncertainty introduced through the modeling of
the $t$-dependence mainly affects the
overall normalization of the cross-section with a mild logarithmic
dependence on the energy. To determine the scattering amplitude, we
first note that the dominant contribution is provided by its imaginary
part. Corrections due to the real part of the scattering amplitude
can be estimated using dispersion relations, in particular 
\begin{align}
  \label{eq:32}
  \frac{\Re\text{e} \mathcal{A}(W^2, t)}{\Im\text{m} \mathcal{A}(W^2, t)}
&=
\tan \frac{\lambda \pi }{2},
& \text{with}&
& \lambda(x) & = 
\frac{d \ln \Im\text{m}  \mathcal{A}(x, t) }{ d \ln 1/x} \, .
%\frac{d \ln \Im\text{m}  \mathcal{A}(W^2, t) }{ d \ln W^2} \, .
\end{align}
As noted in \cite{Bautista:2016xnp}, the dependence of the slope
parameter $\lambda$ on energy $W$ provides a sizable correction to
the  $W$ dependence of the complete
cross-section. We therefore do not assume $\lambda =$const., but instead determine the slope $\lambda$ directly from the $W$-dependent
imaginary part of the scattering amplitude. To determine the latter,
we go beyond the Gaussian model for the light-cone wave function of
the vector mesons and use instead a re-fined description which
includes relativistic spin-rotation effects.  The imaginary part of
the scattering amplitude is then in the forward limit obtained as \cite{Krelina:2018hmt, Cepila:2019skb, Hufner:2000jb}
\begin{align}
  \label{eq:amp}
  \Im\text{m} \mathcal{A}_T(W^2, t=0) & = \int d^2 {\bm r}  \left [\sigma_{q\bar{q}} \left(\frac{M_V^2}{W^2}, r \right)  \overline{\Sigma}_T^{(1)}( r)
 + 
\frac{d \sigma_{q\bar{q}} \left(\frac{M_V^2}{W^2}, r \right)}{dr} \overline{\Sigma}_T^{(2)}(r)
 \right],
\end{align}
with $r = |{\bm r}|$. The functions $\overline{\Sigma}_T^{(1,2)}$
describe the transition of a transverse polarized photon into a vector
meson $V$ and are given by \cite{Cepila:2019skb}
\begin{align}
  \label{eq:sigma}
\overline{\Sigma}_T^{(i)}(r) & = \hat{e}_f \sqrt{\frac{\alpha_{e.m.} N_c}{2 \pi^2}} K_0(m_f r)  \, \Xi^{(i)}(r), \qquad i=1,2 
\end{align}
where
\begin{align}
 \Xi^{(1)}(r)& =  \int\limits_0^1 dz  \int\limits_0^\infty dp\, p J_0(p\cdot r)   \,  \frac{m_T^2 + m_T m_L - 2 { p}^2 z(1-z)}{m_T + m_L} \Psi_V(z, { p}), \notag \\
\Xi^{(2)}(r)& =  \int\limits_0^1 dz  \int\limits_0^\infty dp\, p^2 J_1(p\cdot r) 
  \,  \frac{m_T +  m_L + m_T (1-2z)^2}{2m_T(m_T + m_L)} \Psi_V(z,  p) ,
\end{align}
and $\hat{e}_f = 2/3$ is the charge of the charm quark while $\alpha_{e.m.}$ the electromagnetic fine structure constant; $N_c = 3$ denotes the number of colors and $K_0$ is a Bessel function of the second kind and $J_{0,1}$ a Bessel function of first kind.  Finally, with $m_f$ the mass of the charm quark, and $p = |{\bm p}|$ the modulus of the transverse momentum, we have
\begin{align}
  \label{eq:mT}
  m_T^2 & = m_f^2 + { p}^2    & m_L^2 = 4 m_f^2 z(1-z),
\end{align}
with $\Psi_V(z, { p})$ the wave function of the vector meson. The
latter has been obtained in \cite{Krelina:2018hmt, Cepila:2019skb}
through the numerical solution of the Schr\"odinger equation for a
given choice of the heavy quark interaction potential and provided in
the boosted form as a table in both photon momentum fraction $z$ and
transverse momentum ${ p}$. The above form includes both effects
due to the so-called Melosh spin rotation as well as a more realistic
$r$-dependence of the photon-vector meson transition, with which we
convolute the dipole cross-section $\sigma_{q\bar{q}}(x, r)$. The functions $\overline{\Sigma}^{(i)}$ are plotted against the dipole separation in Fig.~\ref{fig:sigma}. The central observations are the small, but visible node at $r \simeq 0.8$~fm for the $\Psi(2s)$  ($\overline{\Sigma}^{(1)}(r)$) and the relative  enhancement  of  $\Psi(2s)$ with respect to the $J/\Psi$ for   $\overline{\Sigma}^{(2)}(r)$, which is particularly pronounced for the  Harmonic Oscillator potential.  \\

%%%%%%%%%%% gluon
As in \cite{Garcia:2019tne}, we calculate in the following the dipole cross-section from two underlying unintegrated gluon distributions $  {\cal F}(x, {\bm k}^2)$, using the relation \cite{Braun:2000wr} 
\begin{align}
  \label{eq:Nfromugd}
 \sigma_{q\bar{q}}(x, r) & = \frac{4 \pi}{N_c} \int \frac{d^2 {\bm
                           k}}{{\bm k}^2}
\left(1-e^{i {\bm k}\cdot {\bm r}}\right) \alpha_s  {\cal F}(x, {\bm k}^2) \, .
\end{align}
Our study is based on two different implementations, the  KS and HSS unintegrated gluon
 densities respectively: 
\begin{itemize}
\item the KS gluon has been obtained as  a solution of the momentum space version of the  BK equation with modifications
according to the Kwieciński-Martin-Stasto (KMS) prescription \cite{Kwiecinski:1997ee}. This implies an implementation of a so-called  kinematical constraint, leading to energy momentum conservation,  as well as complete DGLAP splitting functions, including quarks. In the collinear limit, the underlying evolution equation  reduces therefore to the conventional DGLAP evolution. The KS gluon distribution in the proton was fitted  \cite{Kutak:2012rf} to proton structure function data  measured at the  HERA experiments H1 and ZEUS \cite{Aaron:2009aa}. For a more detailed discussion see \cite{Kutak:2012rf, Kwiecinski:1997ee}
\item The HSS gluon is subject to NLO BFKL evolution, including a resummation of collinearly enhanced terms in the NLO BFKL kernel as well as a resummation of large running coupling corrections using the optimal scale setting procedure. The initial conditions have been fitted \cite{Hentschinski:2012kr, Hentschinski:2013id} to the same HERA data set as the KS gluon.  For a more detailed discussion see  \cite{Hentschinski:2012kr, Chachamis:2015ona}. 
\end{itemize}
While the HSS gluon provides a very good description of 
 both $\Upsilon$ and $J/\Psi$ photo-production data \cite{Bautista:2016xnp, Garcia:2019tne}, it has been found  in \cite{Garcia:2019tne} that the perturbative expansion used for the solution of the NLO BFKL equation turns unstable at lowest values of $x$. In particular  one finds at the  level of the dipole
cross-section two terms 
\begin{align}
  \label{eq:1}
  \sigma_{q\bar{q}}^{(\text{HSS})}(x, r, M, \overline{M} )  & =  \sigma_{q\bar{q}}^{(\text{dom.})}(x, r, M, \overline{M} )   +   {\sigma}_{q\bar{q}}^{(\text{corr.})}(x, r, M, \overline{M} )   ,
\end{align}
where 
\begin{align}
  \label{eq:2}
\hat{\sigma}_{q\bar{q}}^{(\text{dom})} (x, r, M, \overline{M} )   &= {{\alpha}_s(\overline{M} \cdot Q_0)}%{\bar{\alpha}_s(M^2)}
 \int\limits_{\frac{1}{2} - i \infty} ^{\frac{1}{2} + i \infty}\!\! \frac{d \gamma}{2 \pi i} \left(\frac{4}{r^2 Q_0^2} \right)^\gamma 
 f(\gamma,  \delta, r)   \left(\frac{1}{x}\right)^{\chi\left(\gamma, M, \overline{M} \right)}\notag \\
\hat{\sigma}_{q\bar{q}}^{(\text{corr.})}(x, r, M, \overline{M} )    &=   {{\alpha}_s(\overline{M} \cdot Q_0)} \int\limits_{\frac{1}{2} - i \infty} ^{\frac{1}{2} + i \infty}\!\! \frac{d \gamma}{2 \pi i} \left(\frac{4}{r^2 Q_0^2} \right)^\gamma 
 f(\gamma,  \delta, r)   \left(\frac{1}{x}\right)^{\chi\left(\gamma, M, \overline{M} \right)}\notag \\
&\hspace{-2cm} \times   \frac{\bar{\alpha}_s^2\beta_0  \chi_0 \left(\gamma\right) }{8 N_c} \log{\left(\frac{1}{x}\right)}
  \Bigg[- \psi \left(\delta-\gamma\right)
 +  \log \frac{M^2r^2}{4} - \frac{1}{1-\gamma} - \psi(2-\gamma) - \psi(\gamma) \Bigg]\;,
\end{align}
and
\begin{align}
  \label{eq:5}
  f(\gamma, \delta, r)&= \frac{  r^2 \cdot {\cal C} \pi \Gamma(\gamma) \Gamma(\delta - \gamma) }{ N_c (1-\gamma) \Gamma(2-\gamma) \Gamma(\delta)},
\end{align}
is a function which collects both factors resulting from the proton impact factor and the transformation of the unintegrated gluon density to the dipole cross-section, see \cite{Hentschinski:2012kr, Bautista:2016xnp} for details. The parameters $Q_0 = 0.28$~GeV, $\mathcal{C}=2.29$ and $\delta = 6.5$ have been determined from a fit to  HERA data in \cite{Hentschinski:2012kr}.  $\chi(\gamma,M^2)$ is the next-to-leading
logarithmic (NLL) BFKL kernel which includes a resummation of both collinear enhanced terms as well as  a 
resummation of  large terms proportional to the first coefficient of the QCD
beta function, $\beta_0 =  11 N_c/3 - 2 n_f /3$ 
through the Brodsky-Lepage-Mackenzie (BLM) optimal scale
setting scheme \cite{Brodsky:1982gc},  with $N_c=3$  and $n_f=4$ the number of
colors and active flavors respectively. This procedure yields then in turn a $\gamma$-dependent running coupling constant,  $\bar{\alpha}_s = \alpha^{\text{BLM}}_s(\overline{M} \cdot Q_0, \gamma) N_c/\pi$, see \cite{Hentschinski:2012kr, Hentschinski:2013id} for details.   Running couplings constants are evaluated at $n_f=4$ with $\Lambda_{\text{QCD}} = 0.21$~GeV, see \cite{Hentschinski:2012kr, Hentschinski:2013id} for details. The NLL kernel with collinear
improvements reads
 \begin{align}\label{eq:gluongf}
\chi\left(\gamma, M, \overline{M} \right) &=
{\bar\alpha}_s\chi_0\left(\gamma\right) +
{\bar\alpha}_s^2\tilde{\chi}_1\left(\gamma\right)-\frac{1}{2}{\bar\alpha}_s^2
\chi_0^{\prime}\left(\gamma\right)\chi_0\left(\gamma\right) +
\notag \\
&\hspace{2cm} 
+ \chi_{\text{RG}}({\bar\alpha}_s,\gamma,\tilde{a},\tilde{b}) - \frac{\bar{\alpha}_s^2}{8 N_c} \chi_0(\gamma) \log \frac{\overline{M}^2}{M^2}  .
\end{align}
where $\chi_i$, $i=0,1$ denotes the LO and NLO BFKL eigenvalue and
$\chi_{\text{RG}}$ resums (anti-)collinear poles to all orders; for
details about the individual kernels see
\cite{Hentschinski:2012kr, Bautista:2016xnp}.  The scale $M$ is a
characteristic hard scale of the process, while $\overline{M}$ sets
the scale of the running coupling constant, see
\cite{Hentschinski:2012kr, Bautista:2016xnp} for details.  As in
\cite{Bautista:2016xnp} we consider here the possibility that --
unlike in the original fit -- that $\overline{M} \neq M$ which we use
to estimate the uncertainty in the energy dependence the obtained
dipole cross-section.  The term ${\sigma}^{\text{corr.}}$ contains
running coupling corrections related to the transverse momentum
dependence of external particles which do not exponentiate. They have
been therefore treated in \cite{Hentschinski:2012kr} as a perturbative
correction to the BFKL Green's function.  Even though
${\sigma}^{\text{corr.}}$ is suppressed by a relative factor of
$\alpha_s^2$, enhancement by $\ln(1/x)$ will eventually compensate for
the smallness of the strong coupling constant and invalidate the
perturbative expansion which in turn gives rise to the aforementioned
instability. In \cite{Garcia:2019tne} it has been found that this
instability can be cured through adopting a scale setting, similar to
those used in fits of the so-called IP-sat model
\cite{Rezaeian:2012ji, Bartels:2002cj}, through choosing 
$M^2 = \frac{4}{r^2} + \mu_0^2$ with $ \mu_0^2 =
1.51~\text{GeV}^2$.
It is important to note that this change in the hard scale -- even
though well motivated -- yields a dipole cross-section which does no
longer fit the very precise HERA data; in particular the overall
normalization requires an adjustment. The resulting dipole
distribution provides an opportunity to explore stabilized
perturbative NLO BFKL evolution for the description of exclusive
vector meson photon production, while the parameters $Q_0$ and
$\delta$ could in principle still be further adjusted. We will not
make use of this possibility in this study.  In the following we will
distinguish the two possible implementation of the HSS dipole
cross-section as `fixed' and `dipole' scale respectively.
%\begin{figure}[t]
%  \centering
%  \includegraphics[width=.5\textwidth]{wavefunc2.pdf}
%  \caption{Boosted light-cone wave function of $J/\Psi$ and $\Psi(2s)$ vector mesons respectively. Blue solid and dotted lines show the  $J/\Psi$ wave function at $z=0.5$ and $z=0.7$ while Orange dashed and dot-dashed lines show the $\Psi(2s)$ wave function for  $z=0.5$ and $z=0.7$ respectively. The wave function corresponds to the harmonic oscillator potential of \cite{Krelina:2018hmt,Cepila:2019skb} }
%  \label{fig:wavefuncion}
%\end{figure}'

\begin{figure}[!p]
  \centering
  \includegraphics[width=.7\textwidth]{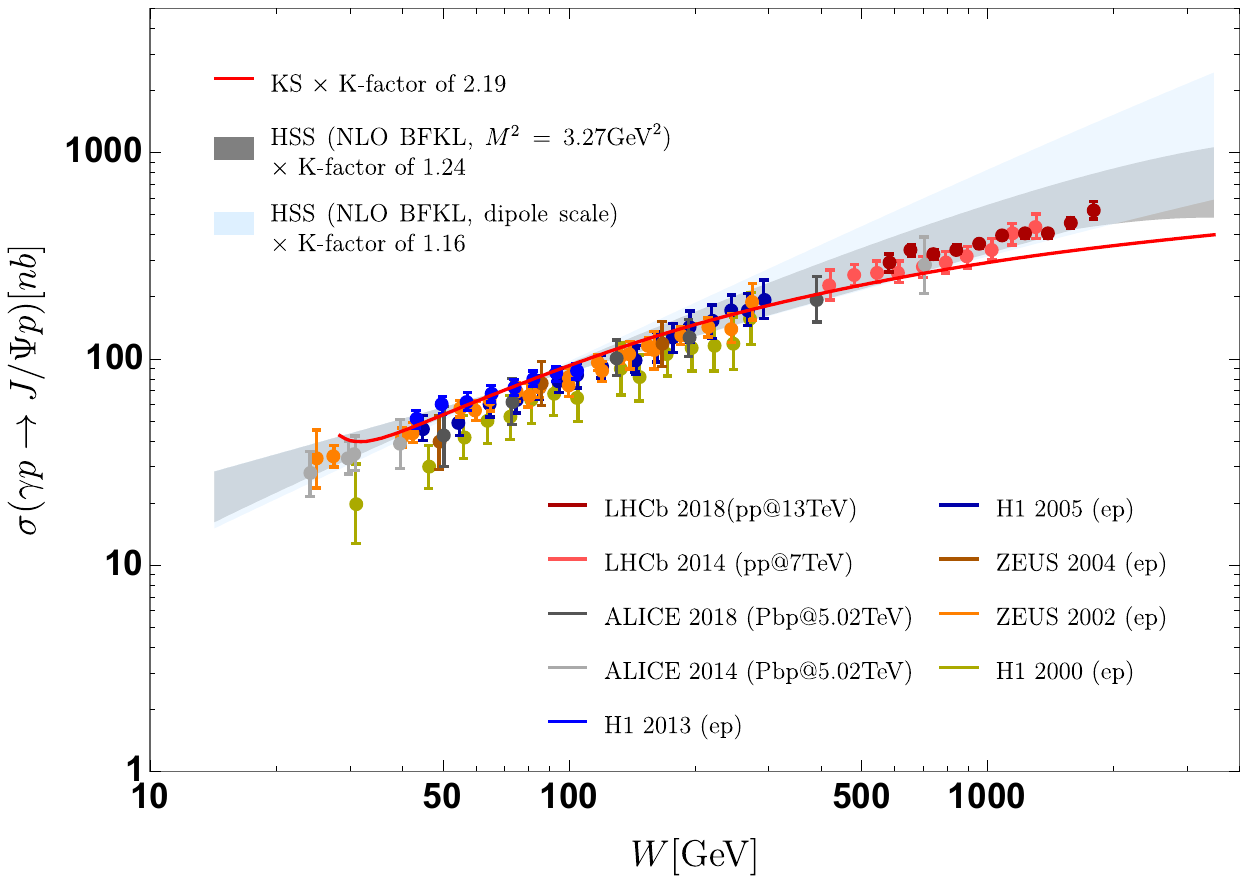}
  \includegraphics[width=.7\textwidth]{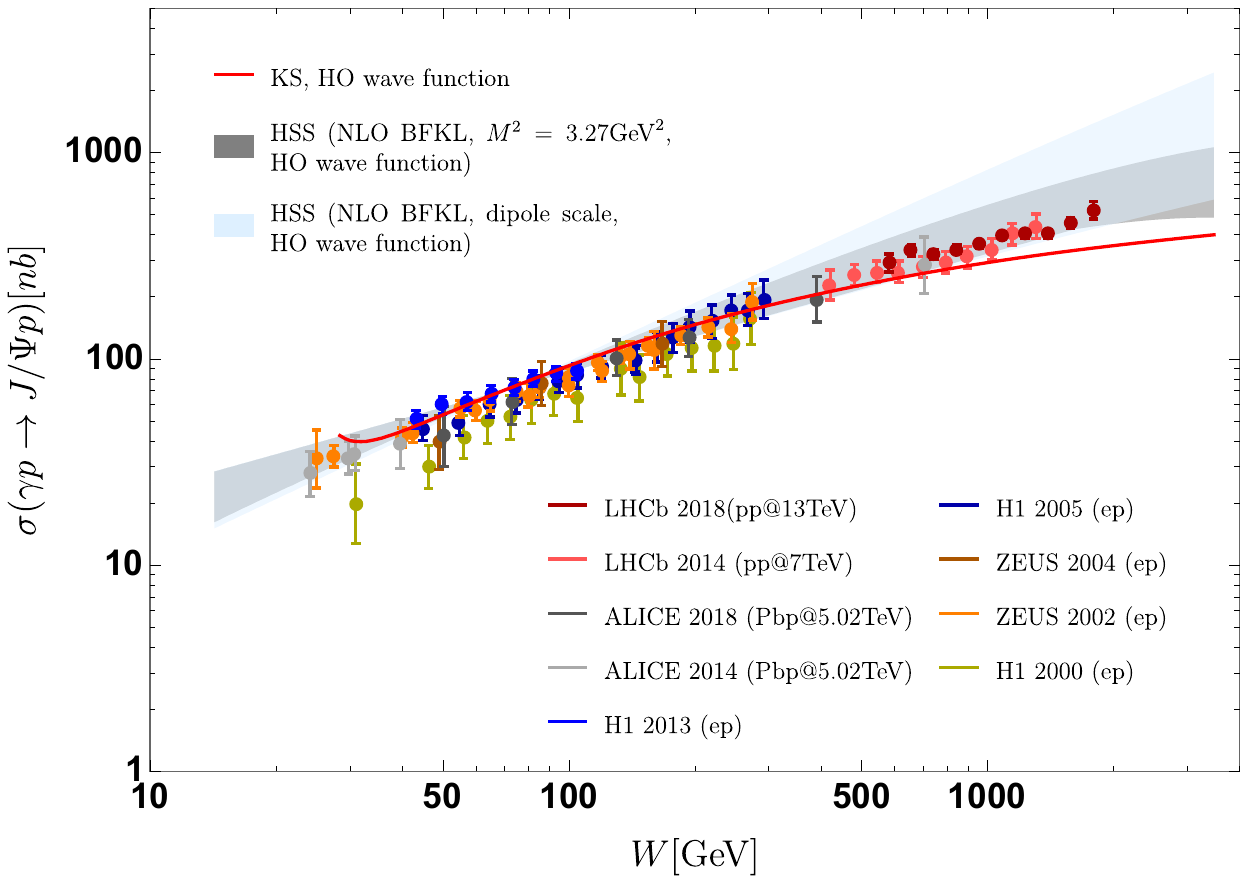}
  \caption{\it Energy dependence of the $J/\Psi$ photo-production
    cross-section as provided by the KS and HSS gluon distribution
    (see text). The shaded regions correspond to a variation of the
    scale $\bar{M} \to \{\bar{M}/{2}, \bar{M} {2}\}$.  The
    upper/lower plot uses $J/\Psi$ wave functions based on the
    Buchm\"uller-Tyle and Harmonic Oscillator potential respectively.
    We further display photo-production data measured at HERA by ZEUS
    \cite{Chekanov:2002xi, Chekanov:2004mw} and H1
    \cite{Alexa:2013xxa,Aktas:2005xu, Adloff:2000vm} as well as LHC data obtained
    from ALICE \cite{TheALICE:2014dwa, Acharya:2018jua} and LHCb
    ($W^+$ solutions) \cite{Aaij:2013jxj, Aaij:2018arx}
    collaborations. }
  \label{fig:results_jpsi}
\end{figure}

\section{Results}
\label{sec:results}

\begin{figure}[!p]
  \centering
  \includegraphics[width=.7\textwidth]{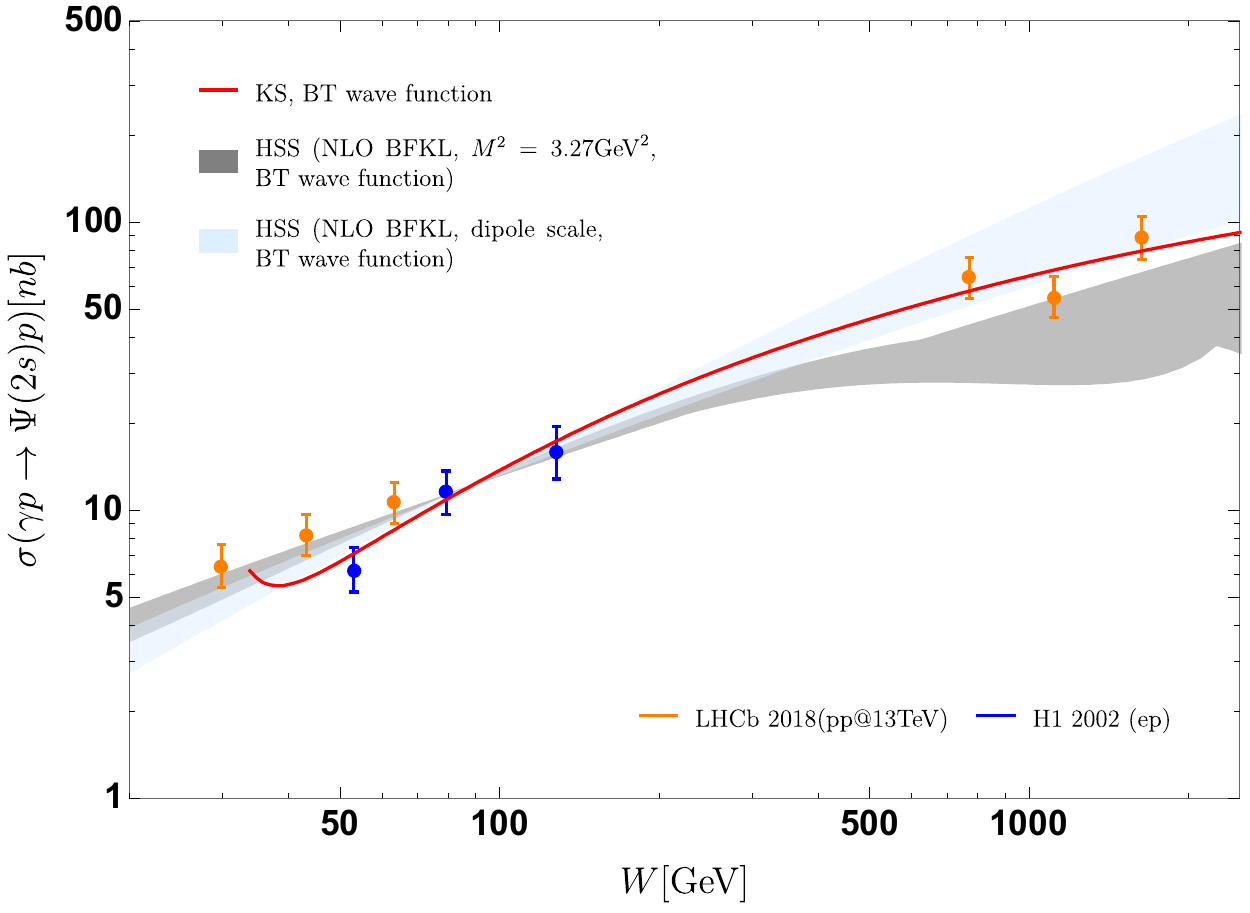}
  \includegraphics[width=.7\textwidth]{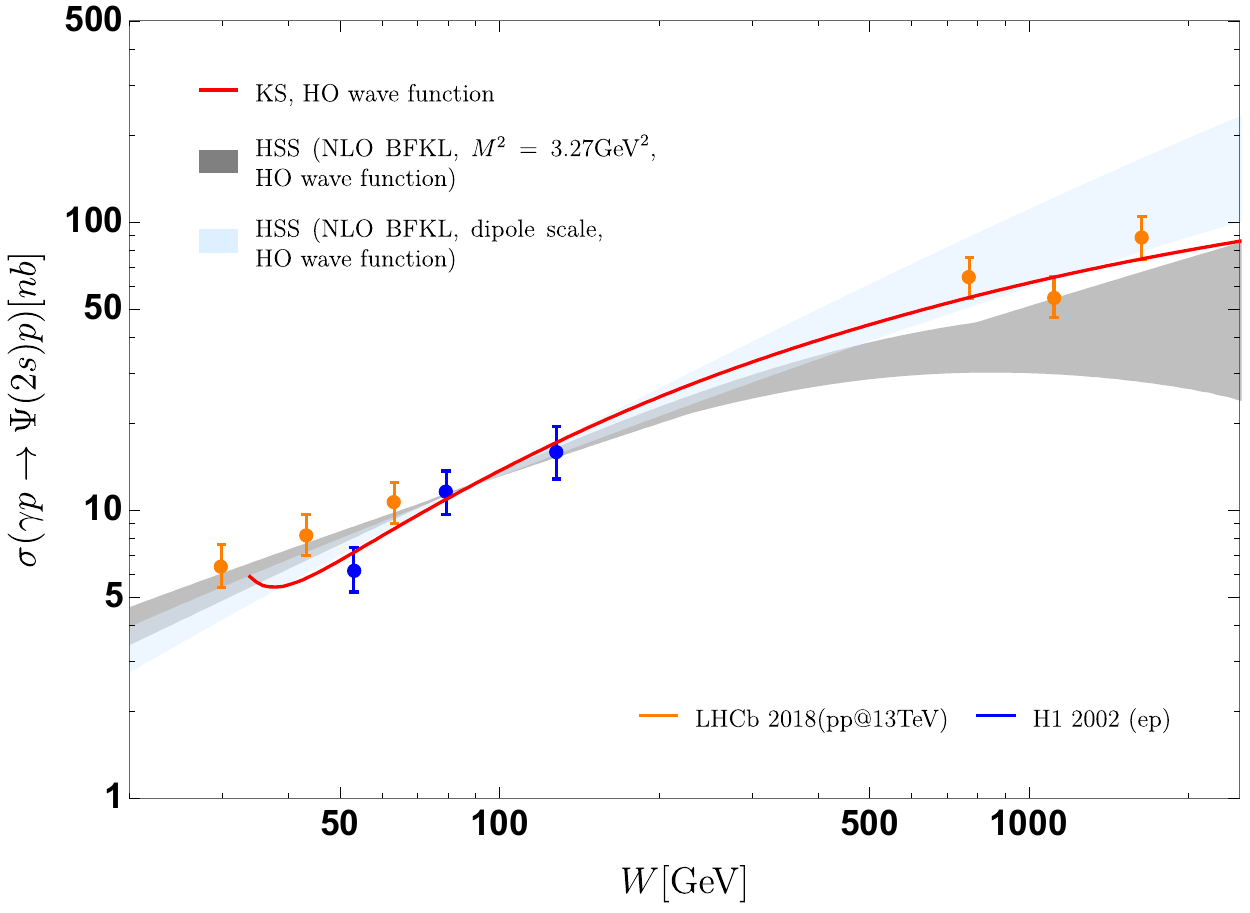}
  \caption{\it Energy dependence of the $\Psi(2s)$ 
    photo-production cross-section as provided by the KS and HSS gluon
    distribution (see text). The shaded regions correspond to a variation of the scale $\bar{M} \to \{\bar{M}/{2}, \bar{M} {2}\}$.  The upper/lower plot uses the wave function based on the the Buchm\"uller-Tyle and  Harmonic Oscillator potential respectively. 
     We
    further display photo-production data measured at HERA by the H1
    \cite{Schmidt:2001tu, Adloff:2002re} as well as LHC data obtained
    from  the  LHCb collaboration ($W^+$ and $W^-$ solutions)
    \cite{ Aaij:2018arx}. }
  \label{fig:results_psi2s}
\end{figure}

For our study we make use of two sets of vector meson wave functions
provided by the authors of \cite{Krelina:2018hmt, Cepila:2019skb}. They
are based on a numerical solution of the Schr\"odinger equation with
harmonic oscillator (HO) and Buchm\"uller-Tye potential
\cite{Buchmuller:1980su} respectively which has been performed and is
provided by the authors of \cite{Krelina:2018hmt, Cepila:2019skb}; we
refer also to these references for a compact summary of the precise
form of the underlying charm-anticharm potentials. While the
Buchm\"uller-Tye potential uses a charm mass of $m_f = 1.48$~GeV, the
harmonic oscillator potential is associated with a charm mass of
$m_f = 1.4$~GeV.For the parameters of the diffractive slope $B_D$,
defined in Eq.~\eqref{eq:18}, we use the following parameters
which have been determined in \cite{Cepila:2019skb} from a fit to HERA data:
\begin{align}
  \label{eq:BD}
  b_0^{(J/\Psi)} & = 4.62, & b_0^{\Psi(2s)} & =  b_0^{(J/\Psi)} + 0.24, \notag \\
 \alpha'_{J/\Psi}(0) & = 0.171, &  \alpha'_{\Psi(2s)}(0)& =  \alpha'_{J/\Psi}(0) - 0.02. 
\end{align}
In agreement with the original fits of the  KS gluon, the
overall strong coupling constant which arises from Eq.~\eqref{eq:1} is
evaluated at the charm mass for the KS gluon with $\alpha_s(m_c) = 0.31$. In the case of the HSS gluon, this coupling constant is evaluated following Eq.~\eqref{eq:2}. The results of our study for the $J/\Psi$
cross-section are shown in Fig.~\ref{fig:results_jpsi}, for the $\Psi(2s)$ in Fig.~\ref{fig:results_psi2s}. To estimate uncertainties associated with the low $x$ evolution,  we vary for the HSS gluon the hard scale in the range $M \to M/2, 2 \cdot M$. For both KS and HSS gluon the following statement applies: Since  the cross-section is proportional to the square of the strong coupling constant, the overall normalization is  strongly dependent on the value of the overall coupling constant. Moreover, due to  the  absence of next-to-leading order corrections for the photon-vector meson impact factor, the value of this coupling is not well constrained and leads to a significant scale uncertainty, and can easily yield changes in the normalization in the range of an overall factor of $0.51$ up to $1.65$. Moreover in  \cite{Krelina:2018hmt, Cepila:2019skb} a strong dependence of the normalization of the photon-vector meson impact factor on the model dependent value of the charm mass has been found. Further uncertainties in the overall normalization arise due to  the absence
of the so-called skewness corrections, see \cite{Shuvaev:1999ce,
  Martin:1999wb}. As pointed out in \cite{Bautista:2016xnp},
we believe that it is not clear whether the approximations used in
\cite{Shuvaev:1999ce, Martin:1999wb}  are appropriate for the current
setup based on high energy factorization, see also the related
discussion in \cite{Cepila:2019skb}.  We therefore do not include a
skewness factor. Nevertheless, the skewness correction of \cite{Shuvaev:1999ce, Martin:1999wb} would yield in our case with the given values for the effective intercept $\lambda$, see Eq.~\eqref{eq:32}, a correction in the overall normalization  which ranges between a factor $0.872$ and  $1.53$. To separate this normalization uncertainty from the uncertainty associated with the description of the energy dependence, on which we focus in this paper, we fit in the following the overall normalization of our theory  prediction to low energy H1 data for $J/\Psi$ photo-production\cite{Alexa:2013xxa} (with $W/$GeV$^2 \in [43.2, 104.2]$)   and the ratio  of  $\Psi(2s)$ and $J/\Psi$ cross-section \cite{Adloff:2002re}  (with $W/$GeV$^2 \in [53.2, 128.3]$). We chose here   the ratio  of  $\Psi(2s)$ and $J/\Psi$ cross-section,  as a reference since to the best of our knowledge these the only  published and independently determined low energy data for  $\Psi(2s)$ photo-production, which are currently available. Note that we do not only make such an adjustment for different versions of the KS and HSS gluon, but we also adjust separately the overall normalization if we vary the renormalization scale of the HSS gluon in the range $M \to M/2, 2M$. In this way our uncertainty bands shows only  the uncertainty in the low $x$ evolution and not in the overall normalization, which is easily twice as big. The shown predictions are therefore obtained through Eq.~\eqref{eq:16total} which is then  multiplied by the adjusted normalization factors,  collected in Tab.~\ref{tab:kfactors}. In some cases  these adjustments in the overall normalization are rather large and can reach values of up to $2.19$. Nevertheless, given the uncertainties in the overall normalization listed above, these values have a natural explanation and are therefore reasonable within the current limitations of the description.
\begin{table}[t]
\centering

\begin{tabular}{*9c}
\toprule
  &  KS (BT)& \multicolumn{3}{c}{HSS (dipole scale, BT)} & \multicolumn{3}{c}{HSS (fixed scale, BT)}  \\
\midrule {}
 && $M/2$ & $M$ & $2 M$ & $M/2$ & $M$ & $2 M$ \\ 
\midrule 
$J/\Psi$ & 2.17 & 1.92 & 1.36 & 1.36 & 1.91 & 1.23 & 1.25 \\
ratio & 0.74 & 0.92 & 1.10 & 1.06 & 0.68 & 0.94 &  0.99 \\
$\Psi(2s)$ & 1.60 & 1.76 & 1.49 & 1.44 & 1.30 & 1.16 & 1.24 \\ 
\midrule
 &  KS (HO)& \multicolumn{3}{c}{HSS (dipole scale, HO)} & \multicolumn{3}{c}{HSS (fixed scale, HO)}  \\
\midrule {}
 && $M/2$ & $M$ & $2 M$ & $M/2$ & $M$ & $2 M$ \\ 
\midrule 
$J/\Psi$ &  2.19 & 1.81 & 1.24 & 1.25 & 1.94 & 1.16 & 1.16\\
ratio &  0.41 & 0.56 & 0.67 & 0.65 & 0.39 & 0.57 & 0.61 \\
$\Psi(2s)$ &  0.92 & 1.01 & 0.83 & 0.81 & 0.74 & 0.66 & 0.71 \\
\bottomrule
\end{tabular}
\caption{Results of the re-fit of the overall normalization to H1 $J/\Psi$ data \cite{Alexa:2013xxa} with $W/$GeV$ \in [43.2, 104.2]$ and H1 data for the $\Psi(2s)$-$J/\Psi$ ratio \cite{Adloff:2002re}  with $W/${GeV}$ \in [53.2, 128.3]$.  Values for $\Psi(2s)$ are calculated as a product of the normalization of ratio and $J/Psi$.} 
  \label{tab:kfactors}
\end{table}
 \\

For the description of the energy dependence of $J/\Psi$ photo-production, Fig.~\ref{fig:results_jpsi}, we confirm the observation made in 
\cite{Garcia:2019tne}: the fixed scale BFKL dipole follows the non-linear KS gluon, which can be explained due to previously mentioned instability of this solution at largest values of $W$. Nevertheless,  the uncertainty band associated with the
dipole scale HSS gluon does no longer allow to clearly discard this
solution through the data. Indeed, the non-linear KS gluon seems to
slightly undershoot the data at highest $W$-values and therefore can
be no longer identified as the preferred description. In
addition, similar to the case of the HSS gluon, one should also
associate with the KS gluon an uncertainty band, which we estimate to
be similar in magnitude or even larger than the one of the HSS
gluon. At the same time it should be stressed that the error bars
shown for the LHCb data at highest values of $W$ reflect only the
error associated with the hadronic cross-sections and uncertainties
due to the extraction of the photon-proton cross-section are not
included. It is therefore likely that these error bars do not reflect
the complete uncertainty associated with these data points. We
therefore conclude that it is not
possible to clearly identify one of the two gluons as the appropriate
description of the currently available $J/\Psi$ data set.
\\

The situation is even less clear if we turn to the $\Psi(2s)$
photo-production cross-section Fig.~\ref{fig:results_psi2s}. While the
dipole scale HSS gluon and the KS gluon both provide a very good
description of the energy dependence with essentially identical
result for the wave function based on Buchm\"uller-Tyle and Harmonic
Oscillator potential, the consequences of the instability of the fixed scale HSS gluon are even more severe in this case. Indeed, starting with $W > 780$~GeV (BT wave function) and  $W >  880$~GeV  some of the solutions explored within the uncertainty band develop a negative intercept $\lambda$, see Eq.~\eqref{eq:32}, and solutions associated with lower and upper limits of the scale variations start to cross. For  $x < 3.50 \cdot 10^{-6}$, corresponding to   $W> 1970$~GeV, we find values  of $\lambda < - 0.5$, and  we clearly leave the region of applicability of Eq.~\eqref{eq:32} to determine the real part through the imaginary part. Indeed, Eq.~\eqref{eq:32}  suggests in this case that the sub-leading real part would be larger (with respect to its absolute value) than the corresponding imaginary part. For  center of mass energies $W > 3.5$~TeV (not shown) one eventually reaches values   $\lambda < - 1$ and the description breaks down completely. A few details on the origin of this instability of the fixed order description in the case of  $\Psi(2s)$ production are collected in the appendix, which explain the origin of this instability both in terms of the perturbative expansion underlying Eq.~\eqref{eq:1} and the particular structure of the $\Psi(2s)$ wave function. \\

\begin{figure}[!p]
  \centering
  \includegraphics[width=.7\textwidth]{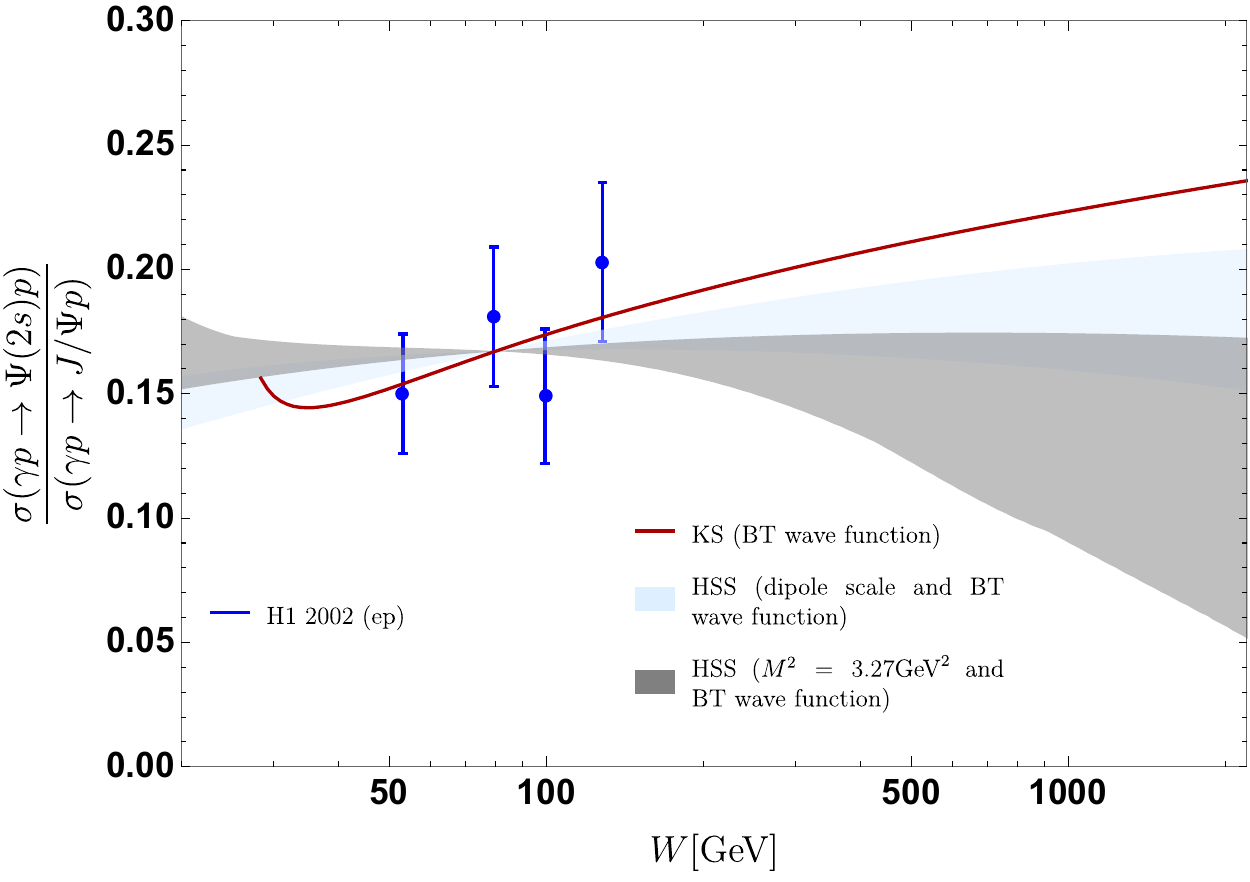}
 \includegraphics[width=.7\textwidth]{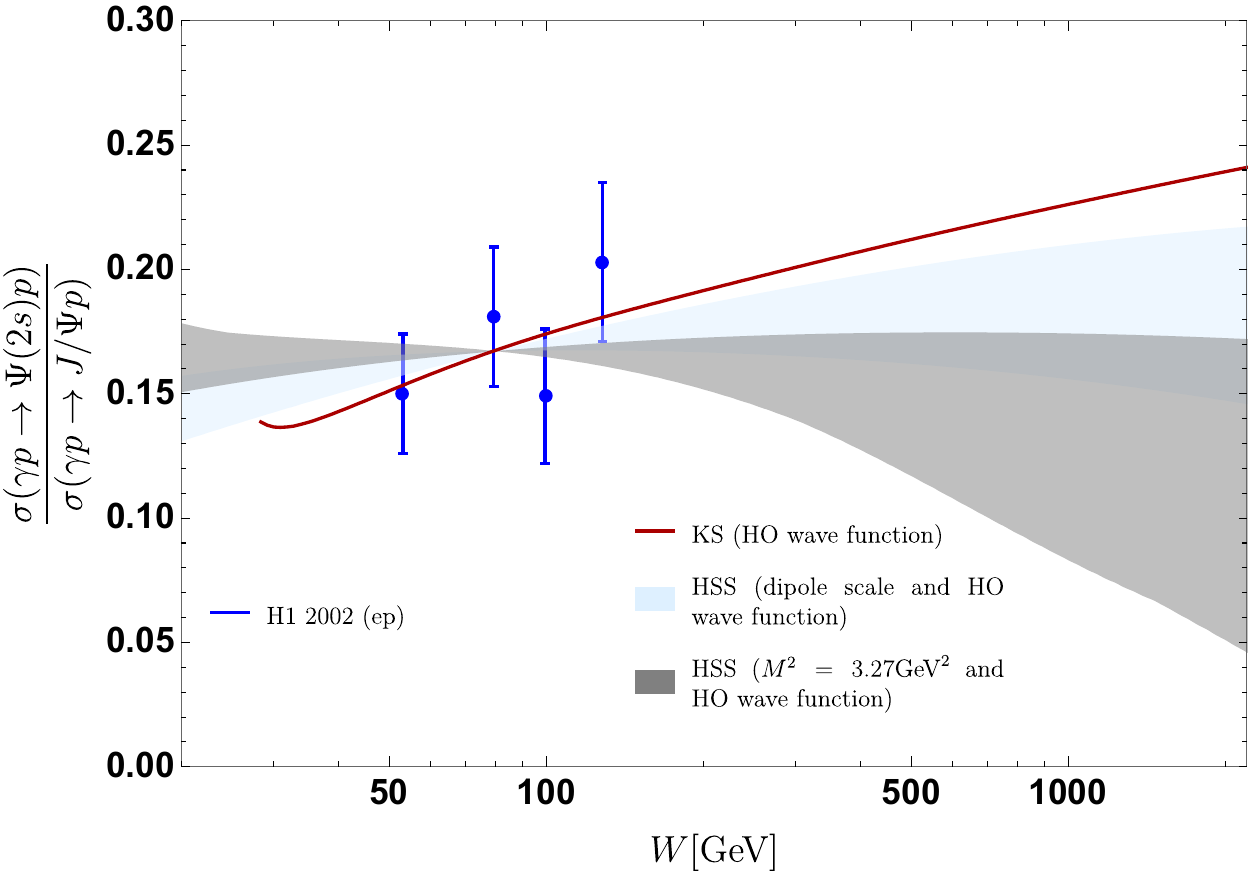}
  \caption{Energy dependence of the ratio of $\Psi(2s)$ vs. $J/\Psi$
    photo-production cross-section as provided by the KS and HSS gluon
    distributions for both Buchm\"uller-Tye (BT) and Harmonic
    Oscillator (HO) vector meson wave functions. The shaded regions
    correspond to a variation of the scale
    $\bar{M} \to \{\bar{M}/2, \bar{M} 2\}$ with the normalization for each scale setting individually fixed through a fit to H1 data, see Tab.~\ref{tab:kfactors}.   We further
    display photo-production data measured at HERA by the H1
    collaboration
    \cite{Adloff:2002re}.} 
  \label{fig:ratio}
\end{figure}
While both the $J/\Psi$ and $\Psi(2s)$ photo-production cross-section
can currently not distinguish between linear and non-linear QCD
evolution, we make an interesting observation if we consider instead
the ratio of both cross-section, Fig.~\ref{fig:ratio}. We find that the KS gluon, subject to
non-linear BK evolution, and the dipole scale HSS gluon, subject to linear
NLO evolution, predict a different energy dependence for the ratio of
$\Psi(2s)$ and $J/\Psi$ photo-production cross-sections. While linear
NLO BFKL evolution predicts a ratio which is approximately constant with
energy, non-linear KS evolution predicts an increase with energy of
the cross-section ratio. While the uncertainty of the HSS fixed scale solution is large and should be considered with care, given the existing problems in the description of the $\Psi(2s)$ photo-production cross-section,   this ratio seems  to decrease with energy in this case. In particular, while the fixed scale HSS gluon and the KS gluon give very similar predictions for the $J/\Psi$ photo-production cross-section, the corresponding predictions are raher different for the ratio.     We believe that this observation can be
useful for two reasons: First of all it is well known that
uncertainties are generally reduced for such cross-section
ratios. This refers both to the aforementioned skewness factor as well
as to the extraction of the photo-production cross-section from hadronic
data, which requires to control the so-called rapidity gap survival
probability. Second, while the differences between linear and
non-linear evolution are in general not large at current
center-of-mass energies, they seem to follow a different tendency,
{\it i.e.} the cross-section ratio increases for non-linear evolution
and decreases for linear evolution. Moreover, linear predictions which mimic the energy dependence of the non-linear gluon through a perturbative instability lead -- at least in the present case -- to a very different energy dependence for the ratio.  As far as data are concerned, we
find that H1 data seem to prefer a rise of the ratio with
energy. Nevertheless, due to the relative large error bars as well as
their limitation to the region $W = 50-110$~GeV, the H1 data set is in
complete agreement with both linear and non-linear evolution.  LHCb
data, which would cover the region of large energies $W$, are
currently only provided for $J/\Psi$ and $\Psi(2s)$ photo-production
cross-sections separately. While it is in principle possible to take
ratios of these results, the published $W$-bins of $J/\Psi$ and
$\Psi(2s)$ cross-sections differ, which complicates a proper
extraction of the cross-section ratio.  We however believe that it
would be very interesting to compare in the future our predictions to
properly extracted cross-section ratios. In particular, regardless of
still size-able theoretical uncertainties, it would be interesting to
see whether experimental data indicate a rising or falling ratio with
energy.

\section{Conclusion}
\label{sec:conlc}

In this paper we extended previous studies, dedicated to the study of
the energy dependence of the exclusive $J/\Psi$ photo-production
cross-section to $\Psi(2s)$ vector mesons. We furthermore used a more
accurate description of the photon to vector meson transition, as
provided by \cite{Krelina:2018hmt, Cepila:2019skb}, as well as a
refined discussion of the theoretical uncertainties of the energy
dependence of the linear HSS gluon.  Reconsidering $J/\Psi$
photo-production including the above mentioned improvements, we find
that we cannot completely confirm the claim made in
\cite{Garcia:2019tne}.  Linear, stabilized HSS evolution, based on the
dipole scale setting and non-linear KS evolution differ for largest
scattering energies $W$, but the difference is not big enough such
that current LHC data can unambiguously distinguish between one of the
two QCD evolution equations, in particular once uncertainties of the
HSS gluon are included. While the difference between HSS gluon with
dipole scale setting and KS gluon is even less pronounced for
$\Psi(2s)$ photo-production, we find that the HSS
gluon with fixed scale suffers a more pronounced instability than observed previously for the $J/\Psi$. While, given the current uncertainties in the theory description,  the energy dependence of $J/\Psi$ and $\Psi(2s)$ seems at current energies not to allow to distinguish between linear and non-linear evolution equations,  we find it encouraging that
the ratio of $J/\Psi$ and $\Psi(2s)$ photo-production cross-section
shows a different energy behavior for linear and non-linear
evolution. In this context we would like to stress that a similar
observation has been already made in \cite{Cepila:2019skb}: with the
gluon modeled through the phenomenological KST dipole cross-section
\cite{Kopeliovich:1999am}, an increase of the ratio with energy has
been found.  At the same time, an almost constant ratio has been found
for the ratio of $\Upsilon(2s)$ and $\Upsilon(1s)$ photo-production
cross-section which are both placed well in the perturbative region
due to the hard scale provided by the bottom quark mass. The current
study goes beyond this observation, since our gluon distributions are
obtained as the solution to low $x$ QCD evolution equations and are both obtained at a hard scale of the order of the charm mass.
\\

From the theory side it is necessary to further increase the accuracy of  predictions for photo-production cross-sections, in particular the rather large adjustment in the overall normalization, see Tab.~\ref{tab:kfactors}. While there are various sources of uncertainty, one may at least expect to reduce the uncertainty in the overall normalization due to a determation of next-to-leading order perturbative corrections to the photon-to-vector meson impact factor, see \cite{Hentschinski:2020tbi, Hentschinski:2014lma, Chachamis:2013hma, Hentschinski:2011tz} for past and recent efforts in this direction. Despite of the theoretical uncertainties, we believe that a precise extraction of the   ratio of $\Psi(2s)$ and  $J/\Psi$  photo-production cross-sections could be very useful  to distinguish in the future between linear and non-linear QCD evolution. We believe that this applies both to photon-proton scattering at highest center of mass energies as measured at LHC, as well as for photo-production cross-sections obtained in electron-ion scattering at  the future Electron Ion Collider. While in the latter case, center-of-mass energies will be naturally lower, nuclear effects will likely enhance gluon densities and therefore the possible relevance of non-linear QCD evolution.

\section*{Acknowledgments}

Support by Consejo Nacional de Ciencia y Tecnolog{\'\i}a grant number
A1 S-43940 (CONACYT-SEP Ciencias B{\'a}sicas) is gratefully
acknowledged. We further would like to thank Andr\'es Nieto Betanzos
for collaboration at an early stage of this project.

\appendix

\section{Details on the BFKL description with fixed scale}
\label{sec:appA}

\begin{figure}[!t]
 % \centering
  \parbox{5cm}{\includegraphics[width=5cm]{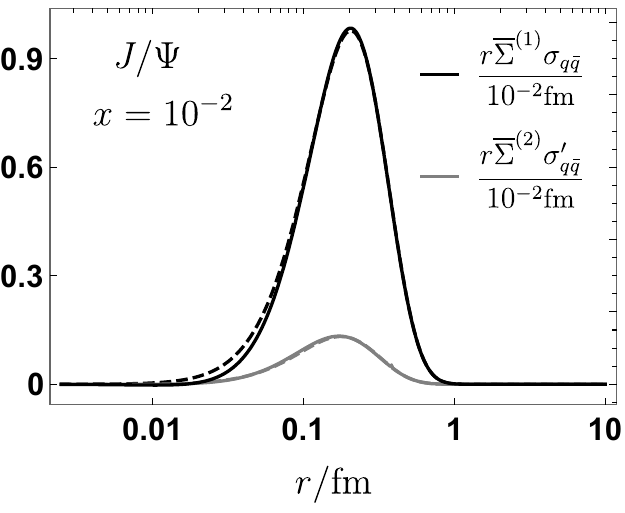}} 
  \parbox{5cm}{\includegraphics[width=5cm]{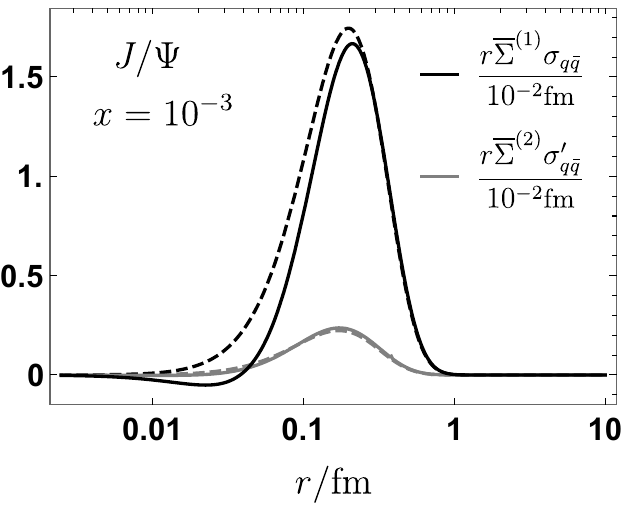}} 
  \parbox{5cm}{\includegraphics[width=5cm]{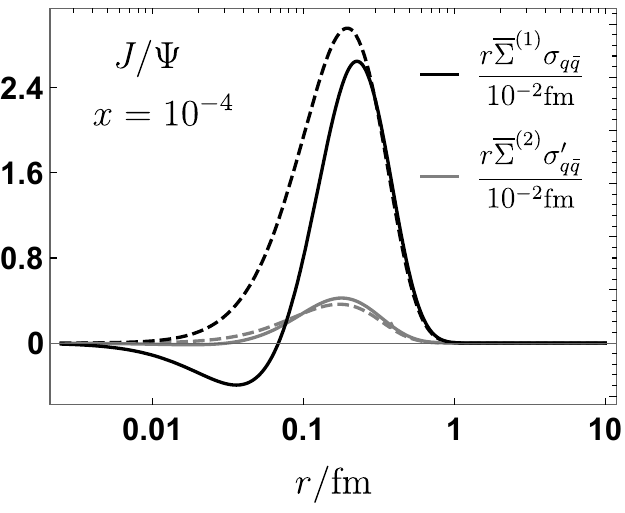}} \\
  \parbox{5cm}{\includegraphics[width=5cm]{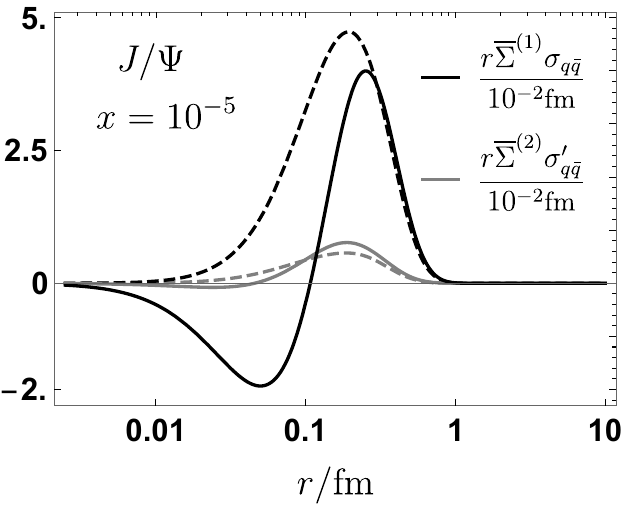}}
  \parbox{5cm}{\includegraphics[width=5cm]{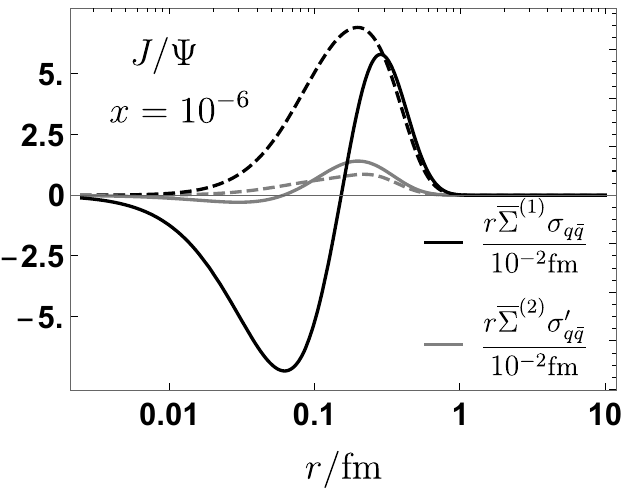}} 
  \parbox{5cm}{\flushright \includegraphics[width=4cm]{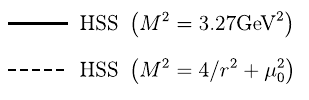}}  
\caption{Overlap of photon to $J/\Psi$ transition light-front wave function and dipole cross-section at different values of $x$. Solid lines correspond to the HSS gluon with fixed external renormalization scale, dashed lines to the dipole size dependent renormalization scale. For this comparison we use the  Buchm\"uller-Tye wave function. }
  \label{fig:fixedVSr2JPsi}
\end{figure}
\begin{figure}[!t]
 % \centering
   \parbox{5cm}{\includegraphics[width=5cm]{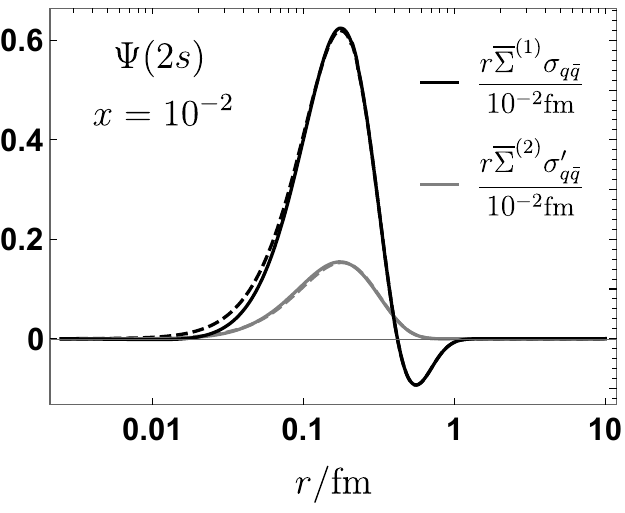}} 
  \parbox{5cm}{\includegraphics[width=5cm]{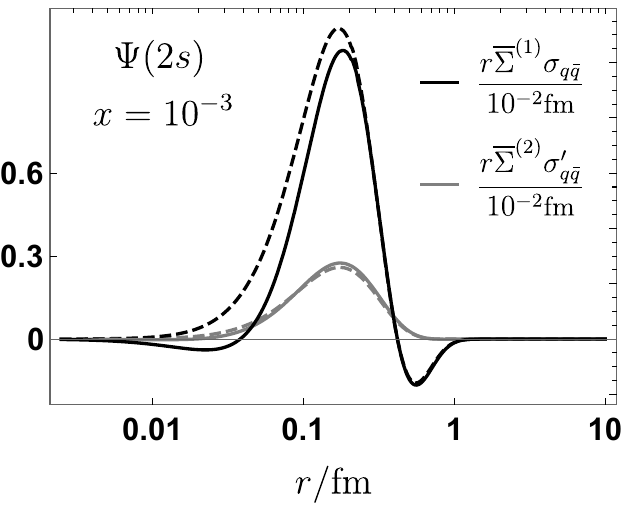}} 
  \parbox{5cm}{\includegraphics[width=5cm]{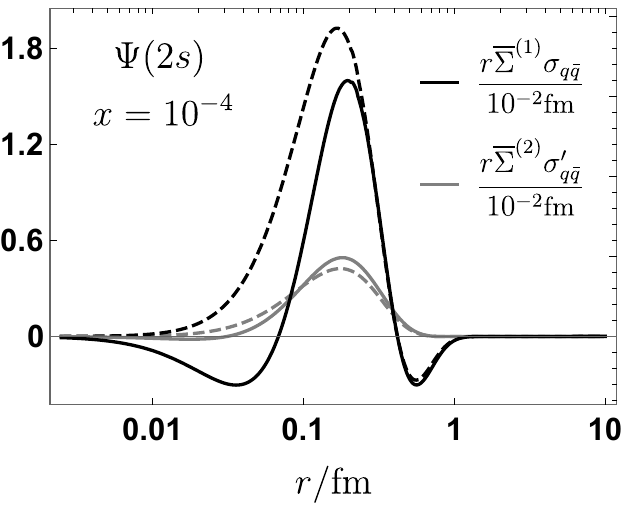}} \\
  \parbox{5cm}{\includegraphics[width=5cm]{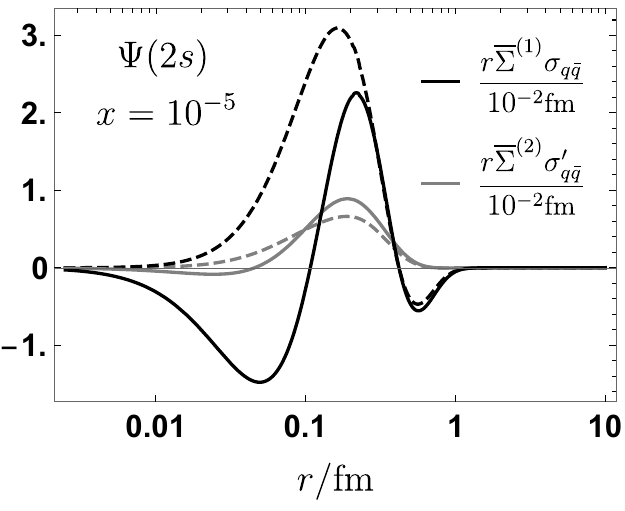}}
  \parbox{5cm}{\includegraphics[width=5cm]{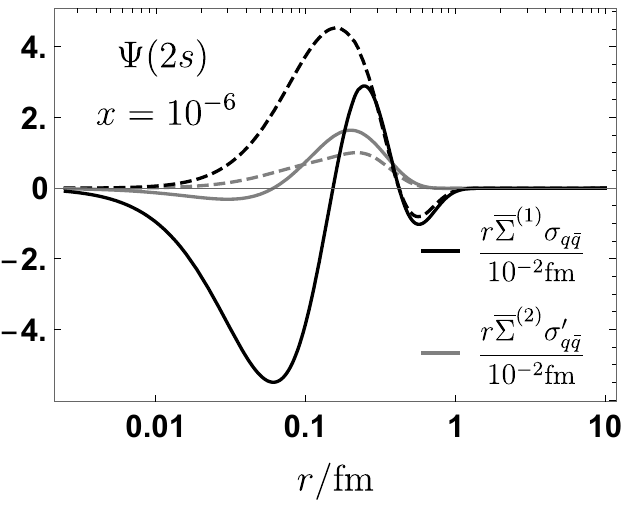}}
 \parbox{5cm}{\flushright \includegraphics[width=4cm]{legend.pdf}}
  \caption{Overlap of the photon to  $\Psi(2s)$ transition light-front wave function and dipole cross-section at different values of $x$. Solid lines correspond to the HSS gluon with fixed external renormalization scale, dashed lines to the dipole size dependent renormalization scale. For this comparison we use the  Buchm\"uller-Tye wave function. }
  \label{fig:fixedVSr2Psi2s}
\end{figure}
In this appendix we provide some details on the instability of the
fixed scale HSS gluon for $\Psi(2s)$ photo-production. As already
pointed out in \cite{Garcia:2019tne} and also discussed in
Sec.~\ref{sec:setup}, the decomposition of the BFKL Green's function
into two terms leads to an instability of the dipole cross-section at
relatively low hard scales and high center of mass energies. Since the
second term in Eq.~\eqref{eq:1} is negative and growing in magnitude
with energy, one finds obtains, due to the presence of a logarithm in
dipole size, a characteristic dip around $r = 0.05$~fm appears, which
leads for $x < 10^{-3}$ to a region of negative dipole cross-sections
which grows with decreasing $x$, see \cite{Garcia:2019tne} for a
detailed discussion. If convoluted with the photon-to-$J/\Psi$ impact
factor this negative region leads in turn to a slow down of the growth
with energy of the scattering amplitude, at least within the range of
energies $W$ accessible at LHC. To illustrate this effect, we provide
in Fig.~\ref{fig:fixedVSr2JPsi} the product of photon-to-$J/\Psi$
transition multiplied with the fixed scale HSS dipole cross-section
against the dipole size for different values of $x = M_V^2/W^2$; for
comparison we further show the dipole size scale HSS case. For lowest
values of $x$, the negative contribution is sizeable in the case of
the fixed scale solution. Nevertheless, after convolution of the HSS
dipole with fixed scale with the $\Sigma^{(1)}$ at a typical low $x$
value of at $x = 0.4 \cdot 10^{-5}$, we still reach 59.8\% of the
corresponding expression obtained with the HSS dipole evaluated at a
dipole size scale.  Albeit the effect of the negative region clearly
affects the theoretical prediction in this region, a description of
data is still possible within the provided uncertainty bands.
\\

In the case of the photon-to-$\Psi(2s)$ transition, one finds two such
effects: a) a negative region in the photon-to-$\Psi(2s)$ transition
at $r \simeq 0.8$~fm due to the presence of a node in the $\Psi(2s)$
wave function, see also Fig.~\ref{fig:sigma}, and the dip region at
$r \simeq 0.05$~fm in the case of the fixed scale HSS solution.  Note
that the presence of this node is of particular importance for
non-saturated gluons, since the latter typical grow with dipole size
$r$. This  leads to an enhancement of this region with respect to a saturated
gluon, which approaches a constant value for large values of $r$.  The
corresponding integrands are shown in
Fig.~\ref{fig:fixedVSr2Psi2s}. While in the case of the dipole size
scale HSS gluon, the negative contribution due to the node at is
compensated by the positive contributions at $r \simeq 0.1$~fm, the
dipole cross-section itself turns negative at such values of $r$ for
the fixed scale HSS gluon. The combination of both effects leads then
to an even stronger reduction of the scattering amplitude with $x$,
since both negative regions increase with energy. As a consequence, the
ratio of fixed scale HSS gluon, convoluted with $\Sigma^{(1)}$, and
the corresponding expression based on the dipole size scale HSS dipole
amounts now only to 14.6\% at $x = 0.4 \cdot 10^{-5}$. Moreover, the
characteristic growth of the BFKL gluon with energy is reversed and
the scattering amplitude starts to decrease with energy, already in
the region of energies $W$ accessible at LHC.

\end{document}